\documentclass[runningheads]{llncs}
\usepackage[T1]{fontenc}
\usepackage{booktabs}
\usepackage{amsmath}
\usepackage{multirow}
\usepackage{graphicx}
\usepackage{caption}
\usepackage{orcidlink}
\usepackage{subcaption}
\usepackage{todonotes}

\usepackage{listings}
\lstnewenvironment{sql}
{\lstset{
    language=SQL,
    basicstyle=\tiny\ttfamily, 
    breaklines=true,
    breakatwhitespace=false,
    showstringspaces=false,
    commentstyle=\color{gray},
    keywordstyle=\color{blue},
    stringstyle=\color{orange},
    morekeywords={REFERENCES, DATETIME, CONSTRAINT, FOREIGN, KEY},
}}
{}

\usepackage{xcolor}
\begin{document}
\title{A Multi-model Approach for Video Data Retrieval in Autonomous Vehicle Development}
%
%
\author{
Jesper Knapp\inst{1}\thanks{These authors contributed equally to this work.}  \orcidlink{https://orcid.org/0009-0009-7695-2212} \and
Klas Moberg\inst{1}* \orcidlink{https://orcid.org/0009-0007-1952-0558} \and
Yuchuan Jin\inst{2}\orcidlink{https://orcid.org/0000-0002-2791-1117} \and
Simin Sun\inst{1}\orcidlink{https://orcid.org/0009-0003-0873-5968} \and
Miroslaw Staron\inst{1}\orcidlink{https://orcid.org/0000-0002-9052-0864}}
\authorrunning{J.Knapp et al.}
\titlerunning{A Multi-model Approach for Video Data Retrieval}
%
\institute{Chalmers $|$ University of Gothenburg, 417 56 Gothenburg, Sweden \\
\email{\{knapp, klasmob\}@student.chalmers.se, \{simin.su, miroslaw.staron\}@gu.se} \and
Zenseact, Lindholmspiren 2, 417 56 Gothenburg, Sweden\\
\email{yuchuan.jin@zenseact.com}}

\maketitle              
        \begin{abstract}
Autonomous driving software generates enormous amounts of data every second, which software development organizations save for future analysis and testing in the form of logs. However, given the vast size of this data, locating specific scenarios within a collection of vehicle logs can be challenging. Writing the correct SQL queries to find these scenarios requires engineers to have a strong background in SQL and the specific databases in question, further complicating the search process. 
This paper presents and evaluates a pipeline that allows searching for specific scenarios in log collections using natural language descriptions instead of SQL. The generated descriptions were evaluated by engineers working with vehicle logs at the Zenseact on a scale from 1 to 5. Our approach achieved a mean score of 3.3, demonstrating the potential of using a multi-model architecture to improve the software development workflow. We also present an interface that can visualize the query process and visualize the results.

\end{abstract}
\section{Introduction}
The amount of software in modern cars is continually growing and is measured in hundreds of millions of lines of code that process petabytes of data. One of the major drivers of this development in the automotive industry is the autonomous driving functionality. As software advances in autonomous capability, moving from features such as adaptive cruise control and assisted parking toward complete autonomy, the need for increasingly sophisticated software increases \cite{staron2021automotive}.

The many systems required to make autonomous vehicles a reality, generate a large amount of data that needs to be processed and stored at a high rate \cite{sharma2021big}. As autonomous vehicles evolve, this data generation will continue to accelerate. Modern high-end vehicles are equipped with multiple imaging sensors for tasks like object recognition. However, autonomous vehicles usually require diverse types of sensors: cameras, LiDAR, and radar. In addition, autonomous vehicles will need higher resolution images to ensure precision \cite{hussain2018autonomous}. According to an article by IBM, cameras alone in Nvidia testing vehicles generate 1 Terabyte of data per hour \cite{ziegler2020ki}.

This project was carried out in partnership with Zenseact, whose main goal is to develop autonomous software for Volvo Cars. The fleet of test vehicles at Zenseact intended for data collection come equipped with numerous sensors and cameras. All of these sensors and cameras generate high-frequency data that are then processed and combined to form a large set of signals that deliver information to several key systems in the car, such as automatic emergency braking and adaptive cruise control. This set of signals is organized in a large table for each individual test drive with a vehicle.

The data gathered from the vehicles can be very valuable for further development of the software, developers at Zenseact can use historical data to, for example, test edge cases of new features they are implementing. In order to do this, that particular edge case, that we call a scenario, has to be extracted from the corresponding table of signals, one example of such a scenario could be when the vehicle is entering a tunnel.

\begin{figure}[htbp]
    \begin{sql}
    SELECT * FROM
    (
        SELECT 
            d.vehicle,
            d.utc_start,
            d.utc_end,
            d.timestamp,
            d."zen_qm_mapengine/satellite_data/data/longposn/nanodegrees/value"/POWER(10,9) as lon,
            d."zen_qm_mapengine/satellite_data/data/latposn/nanodegrees/value"/POWER(10,9) as lat,
            d."zen_qm_feature_a/zen_qm_feature_a_vehicle_motion_state_data/data/longitudinal_velocity/velocity/meters_per_second/value" as lon_v,
            d."zen_qm_sensorfusion_a/zen_qm_sensorfusion_a_vehicle_motion_state_data/data/lateral_velocity/velocity/meters_per_second/value" as lat_v,
            d."zen_qm_feature_a/lss_diagnostics/data/lka/intervention_info/side" as lka_intervention_side,
            LAG(lka_intervention_side) OVER (PARTITION BY d.vehicle, d.utc_start ORDER BY d.timestamp) as previous_lka_intervention_side,
            d."zen_qm_feature_a/lss_diagnostics/data/lka/status/enable/status_emergency/left/unitless/value" as emergency_enabled,
            d."zen_qm_feature_a/lss_diagnostics/data/lka/status/enable/status/left/unitless/value" as enabled,
        FROM
            read_parquet('parquet_files/{firezone_to_query}/{resim_version}/data/*.parquet.gzip') as d
    ) AS tmp
    WHERE
        (lka_intervention_side = 1 and previous_lka_intervention_side = 0 and enabled = 1 and emergency_enabled = 1)
    \end{sql}
    \caption{SQL code for "Find where left side emergency lka interventions are triggered"}
    \label{fig:sql}
\end{figure}

Existing solutions for finding a particular scenario are often based on querying a database using SQL.  
For instance, consider the query "Find where left side emergency LKA (Lane Keeping Assist) interventions are triggered." The example SQL for this query is shown in Figure \ref{fig:sql}. Note that this query has only one constraint. Now, imagine a more complex search, such as "the vehicle was driving above 15 m/s and about to enter a tunnel with snow on the road in Austria." The SQL for such a query would become significantly more complex. To find a specific scenario, the user needs domain-specific knowledge of the vehicle data and expertise in SQL, making the query creation process very time-consuming. This process is not only costly but also limited by the fact that certain information, such as wind or fog conditions, is not included in the signal data and cannot be queried through SQL alone. 

This project aims to address the challenges that arise when searching for scenarios by exploring and implementing a solution using Generative AI to enable searching for scenarios with natural language, thus avoiding the process of writing complex SQL queries. This can be accomplished by leveraging LLMs to generate descriptions of the scenarios given signal and image data. These descriptions can then be embedded and stored in a vector database, where they can be semantically queried for using natural language queries.  

We aim to address the challenges that arise when searching for scenarios and answer the following research questions:
\begin{itemize}
    \item \textbf{RQ1:} To what extent can an LLM generate a description of a scenario based on signals in the form of tabular data?
    \item \textbf{RQ2:} To what extent can a multi-modal approach, using both signals and video information, enhance the description of driving scenarios?
    \item \textbf{RQ3:} To what degree can the combined descriptions to capture the semantics of a vehicle scenario in a vector representation?
\end{itemize}

The objective of the work in this project is to use multi-model to query the vehicle log database for a specific scenario using natural language. Streamlining the process of searching and understanding vehicle logs by merging information from both signals and images.
\newline

The main contributions of our work is:
\begin{itemize}
    \item Propose using natural language for data retrieval instead of traditional SQL, and confirm its feasibility using a pre-trained model on a small-scale dataset.
    \item Enhance query accuracy by integrating information from multiple input sources.
    \item Investigate an approach that could save precious time for software developers and also enable enable employees without domain specific knowledge to gain insight in the data.
\end{itemize}

\section{Related Work}
The number of studies on the use of LLMs in code generation, summarization and testing is constantly growing (over 1,000 studies published in only 2023 \cite{fan2023large}). Many different implementations of table-to-text generation have been explored previously. One example of table-to-text generation is the generation of biographies using Wikipedia information tables, where the authors trained a language model to output a biography from an input table of information for the person \cite{lebret2016neural}. The technique used to extract information from each field of the table and generate a coherent description is similar to our task since we are extracting data from columns corresponding to specific signals.

Zhao et al. \cite{zhao2023investigating}, explores how well different existing models can answer questions based on information in tabular data, including LLaMA2, GPT-3.5, and GPT-4. The experiments, conducted with several table datasets, show that only GPT-3.5 and GPT-4 could generate reasonable results without a few-shot prompt, where the prompt to the model includes examples of the task. Furthermore, the results show that for some datasets, small models have a comparative performance to large models in understanding of the tables. Since we are limited to using smaller models in this paper, this result is promising.

Similarly, Carballo et al. \cite{carballo2022tabtext} aims to develop an LLM-based approach to generate descriptive text explaining the content of a table in natural language. The generated text can be used in subsequent LLM applications or serve as a means of facilitating comprehension for the readers.

A more recent Microsoft article presents \emph{TableGPT}, an LLM that has been trained to complete a variety of tasks in structured data, one of which is to summarize the contents of a table. In their approach, they use instruction fine-tuning to improve the performance of GPT-3.5 on tabular input, resulting in a substantial improvement in table tasks such as column extraction and error detection \cite{li2023table}. The problem with their approach to tuning the LLM is that the entire table has to fit in the instruction prompt, something that is not feasible for our application since the prompt then would have to fit tokens in the order of billions. This fact is also true for the earlier reviewed paper, where different models were compared; the approach hinges on that the whole table can be fit inside the prompt. However, if the goal is a tool that has a real-world impact, a model in which the table must fit the prompt is rarely enough. The work of analyzing tables often focus on tables that are very large, so the usefulness of such a model would be greatly improved if larger tables could be processed.

The limitation of the previous studies discussed in this section is that the proposed solutions, as mentioned above, are suitable only for tables of a smaller scale compared to the data in this paper. Additionally, a distinction lies in the fact that the models in those studies are trained on tables that include both text and numerical information, whereas this paper will concentrate solely on numerical data.

The concept of using vehicle-gathered data to explain and label scenarios is not entirely groundbreaking. Existing work, exemplified by \emph{Automatic Labeling to Generate Training Data for Online LiDAR-based Moving Object Segmentation} \cite{chen2022automatic}, focuses on harnessing data from test drives-specifically LiDAR signals-to produce labels for training data, for which they reported successful results. In this context, the labels take the form of images with marked points of interest. Although this approach diverges from the methodology and outcomes of the present paper, which revolves around generating textual descriptions using LLMs and vision models, there is a shared objective. Both endeavors aim to generate labels that facilitate future machine learning applications, underscoring the industry's demand for the automated generation of reliable labels. The difference being that this paper aims to generate more broadly applicable labels in the form of textual descriptions instead of specifically labeled images.

There exist partial work in this area, although in other corners of software engineering. For example, the work of Chen et al. \cite{chen2023automated} who used GPT3.5 and GPT4 to generate classes and attributes through prompting to achieve accuracy scores of 0.76 for generating classes, 0.61 for attributes generation, and 0.34 for relationship generation. Despite the use of zero-shot, few-shot and chain-of-thought prompts for these models, the authors admit that more work needs to be done to incorporate the domain knowledge in a correct way. The work of Pandey et al. \cite{pandey2023transdpr} shows that we can capture such higher level concepts, though not with the perfect results yet. Therefore, in this work, we utilize these recent advancements and apply them in the area of scenario identification in autonomous driving. 

\section{Research Design}

We address the proposed research questions by conducting an design science research project with our industrial partner Zenseact \cite{staron2020action}. We combined empirical evaluation with the stakeholders with artefact design and computational experiments to evaluate it.  

\subsection{Dataset}

The data used in this paper are collected from test drives conducted by Zenseact. 

\textbf{Signals:}
During test drives, all sensors and cameras generate high-frequency data, which is processed and combined into a comprehensive set of signals. These signals provide crucial information to several key systems within the car. Each drive's data is saved as a log file and can be accessed as a table in parquet format. These tables can contain up to 6,000 columns, each representing a processed signal. Examples of these signals include detecting snow in the lane ahead or determining if the vehicle is currently inside a tunnel.

Each log corresponds to a single test drive, with durations ranging from less than an hour to a full eight-hour day. The data table captures the entire duration of the drive, resulting in tables with thousands of signals and hundreds of thousands of data points per signal, often totaling on the order of $10^8$ entries. To manage this vast amount of data, logs are divided into \emph{scenarios}, each representing a 30-second segment of the drive. This segmentation reduces the size of each scenario's table to the order of $10^6$ entries, though this is still too large for traditional methods of feeding all the data to a LLM.

\begin{figure}[!htb]
    \centering
    \includegraphics[width=0.6\linewidth]{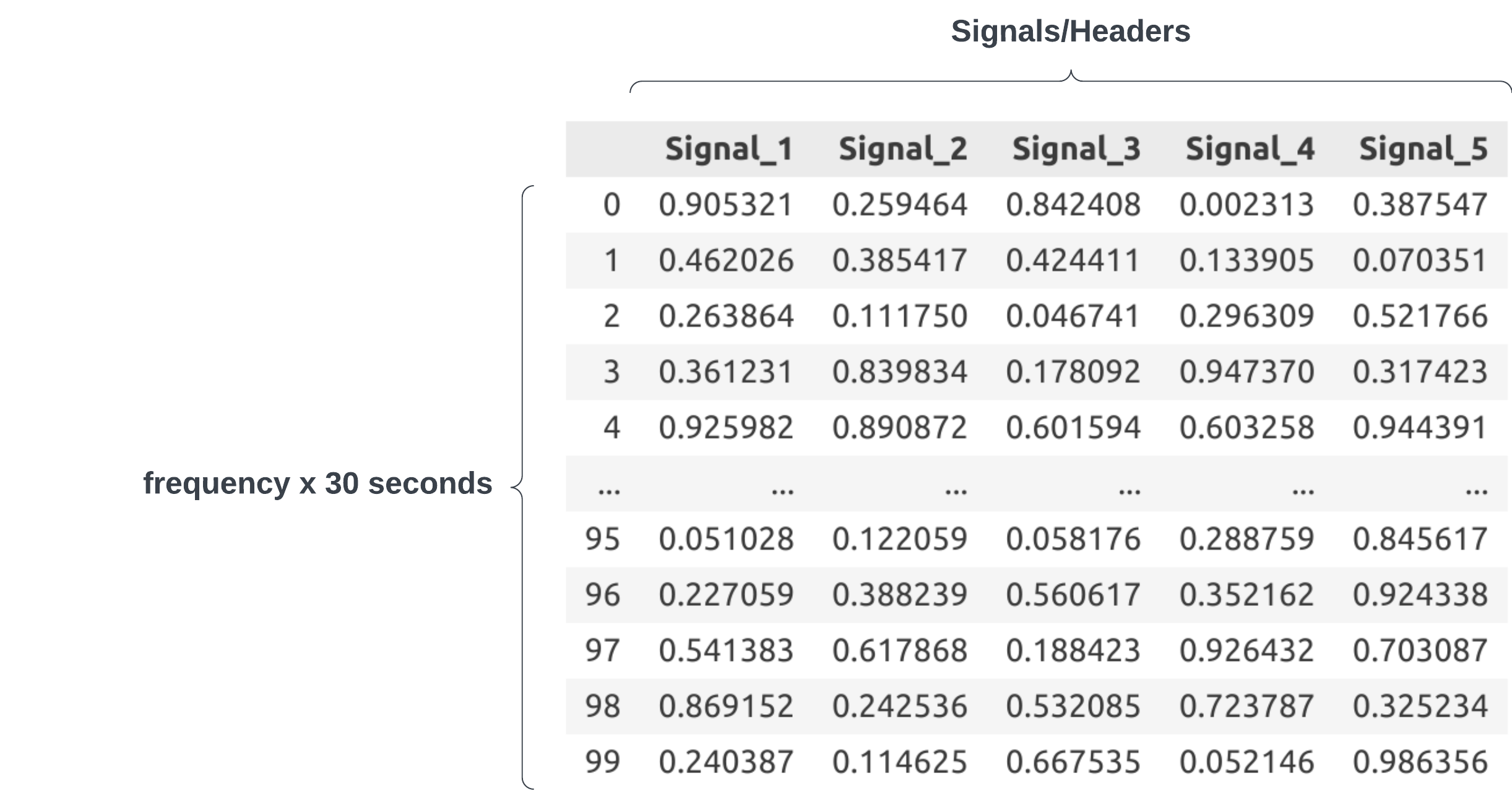}
    \caption{Illustration of vehicle log format with randomized data-points.}
    \label{fig:dsr_cycle}
\end{figure}

\begin{figure}
    \centering
    \begin{minipage}[t]{0.32\textwidth}
        \includegraphics[width=\textwidth]{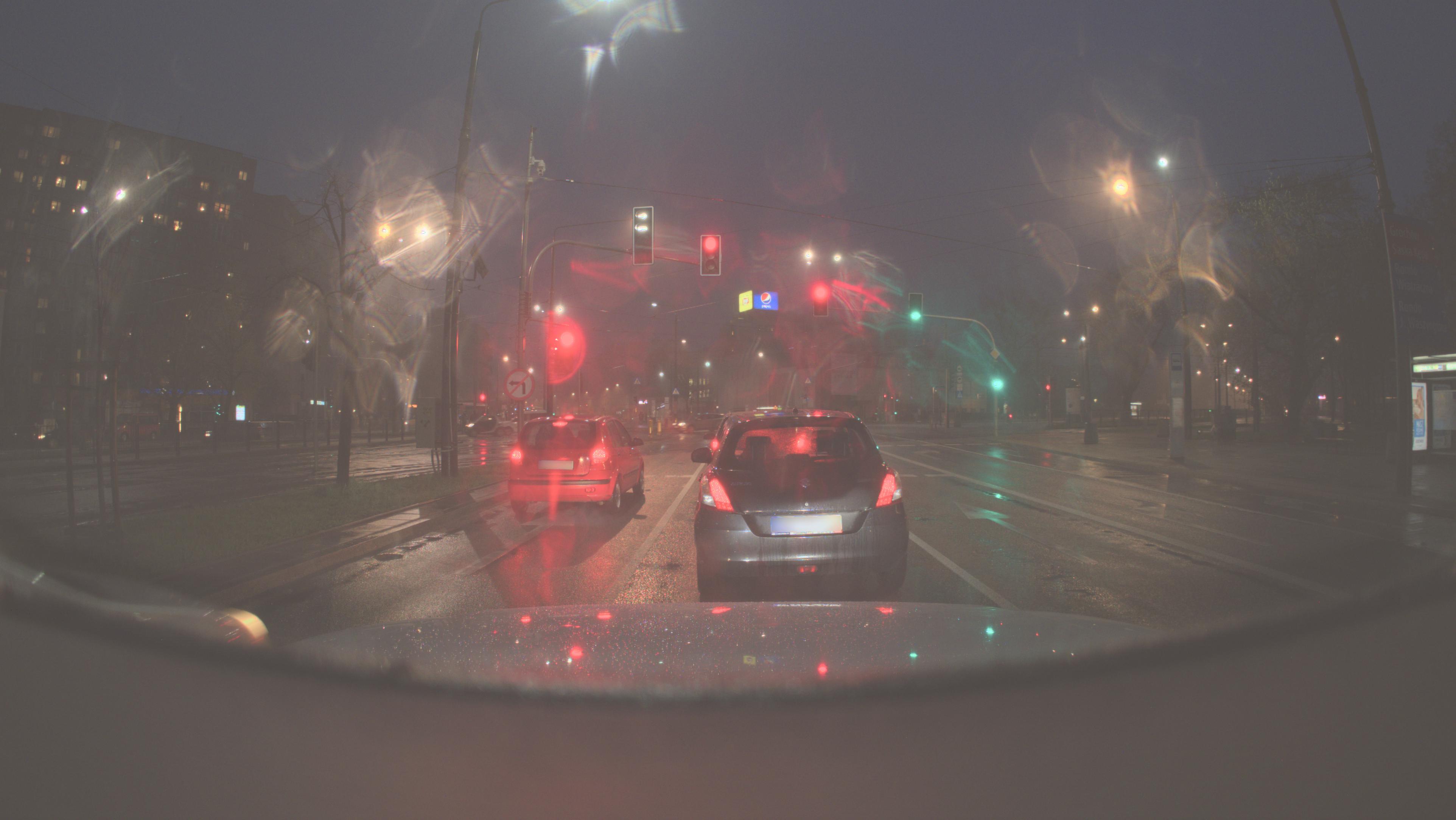}
    \end{minipage}
    \hfill
    \begin{minipage}[t]{0.32\textwidth}
        \includegraphics[width=\textwidth]{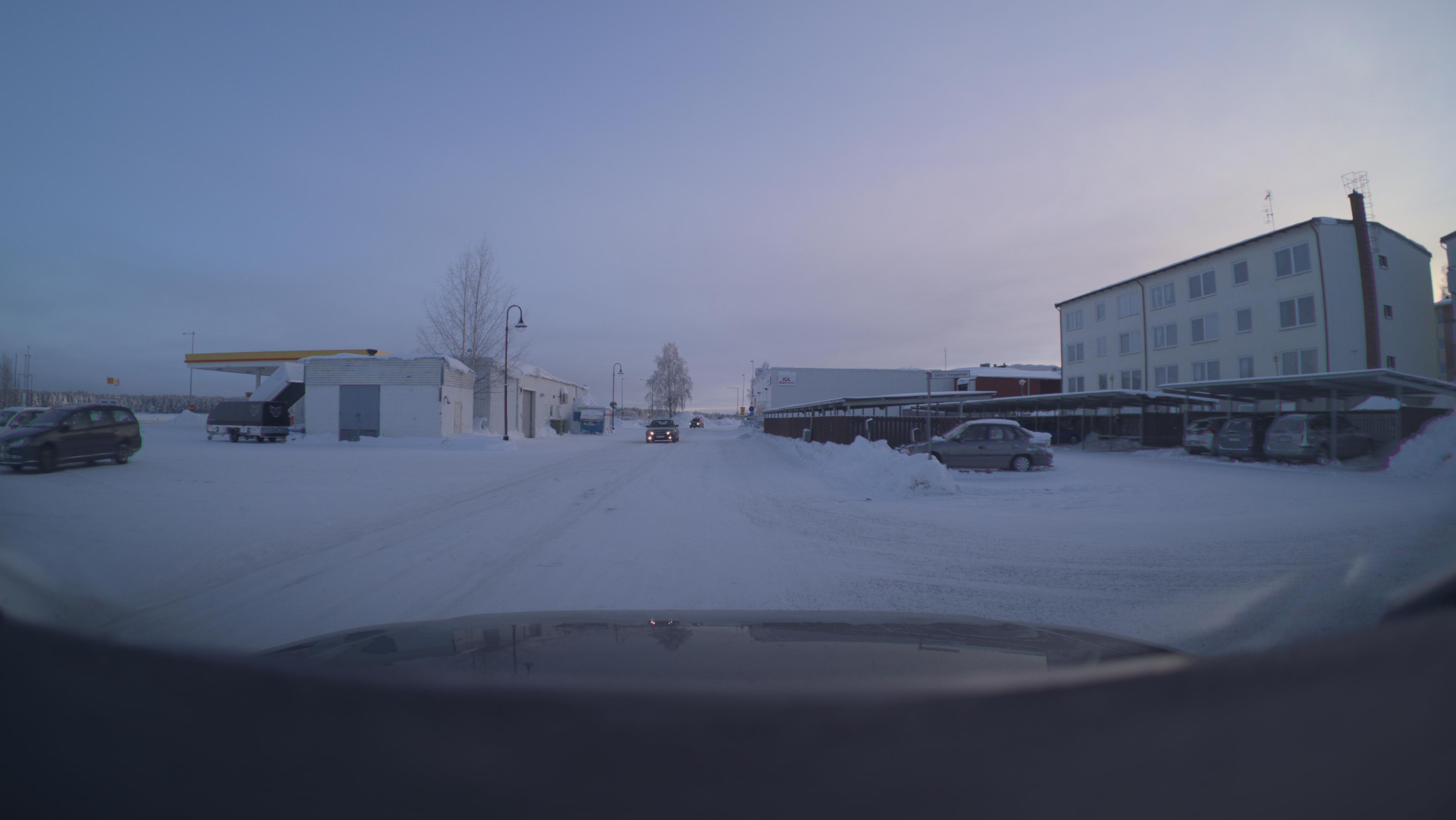}
    \end{minipage}
    \hfill
    \begin{minipage}[t]{0.32\textwidth}
        \includegraphics[width=\textwidth]{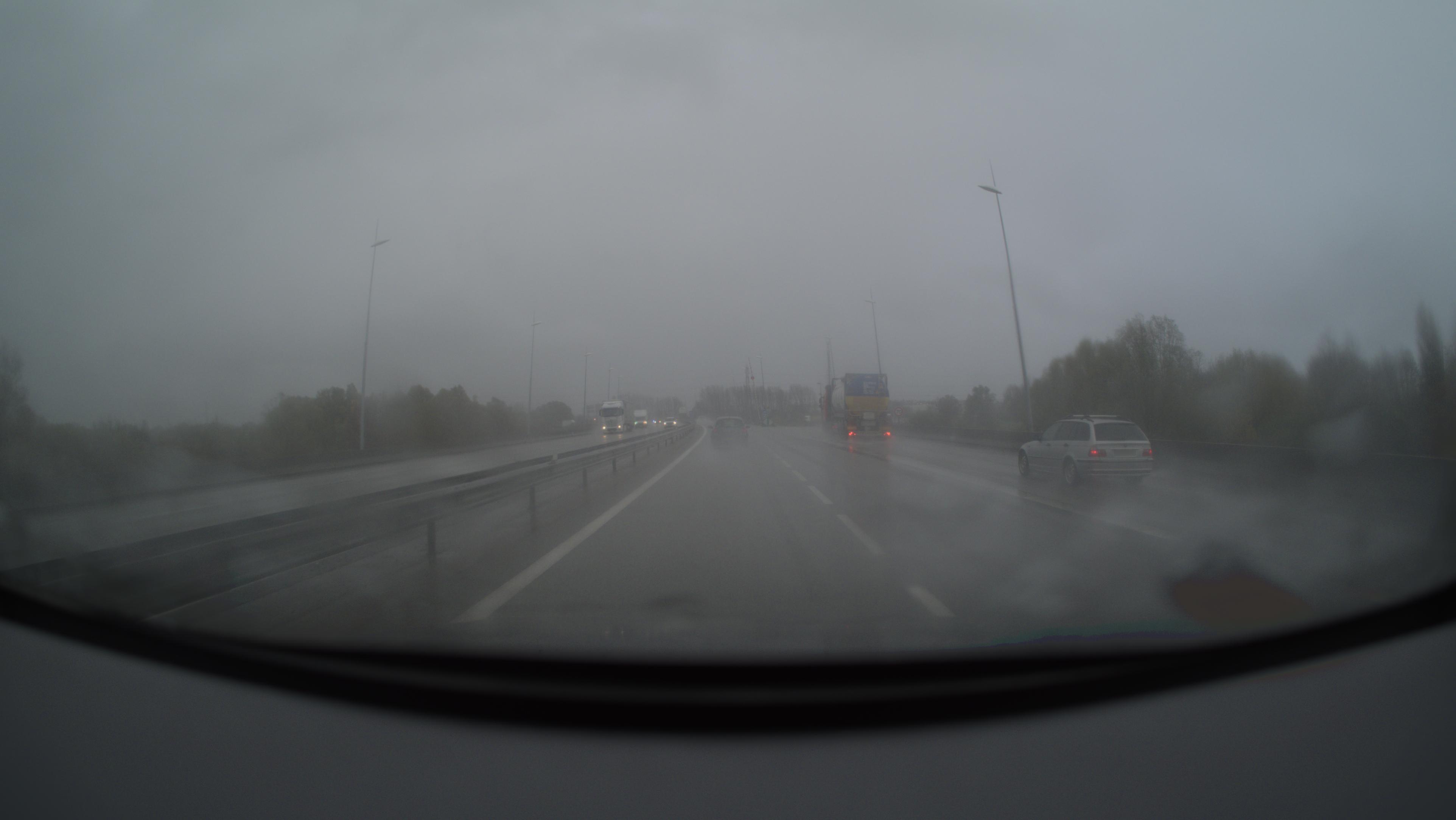}
    \end{minipage}
    \caption{Sample images captured in the vehicle from the Zenseact Open Dataset \cite{zod} }
    \label{fig:data}
\end{figure}

\textbf{Videos:}
The signal data is complemented with video footage captured from a forward-facing camera mounted on top of the vehicle. To utilize both data sources and match the corresponding signals \emph{scenarios}, these videos are divided into 30-second segments and saved in mp4 format. Sample images from those videos are shown in Figure \ref{fig:data}.

Each scenario now consists of a matched pair of a signal snippet and a video snippet. One record represents a single scenario, coupled with additional information such as the link and timestamp. This extra information helps identify the record after performing a search. Examples of vehicle scenarios explored in this paper include driving in a specific country, navigating through a tunnel, and traveling in snowy conditions. 

\subsection{Architecture}
\begin{figure}[!h]
    \includegraphics[width=12cm]{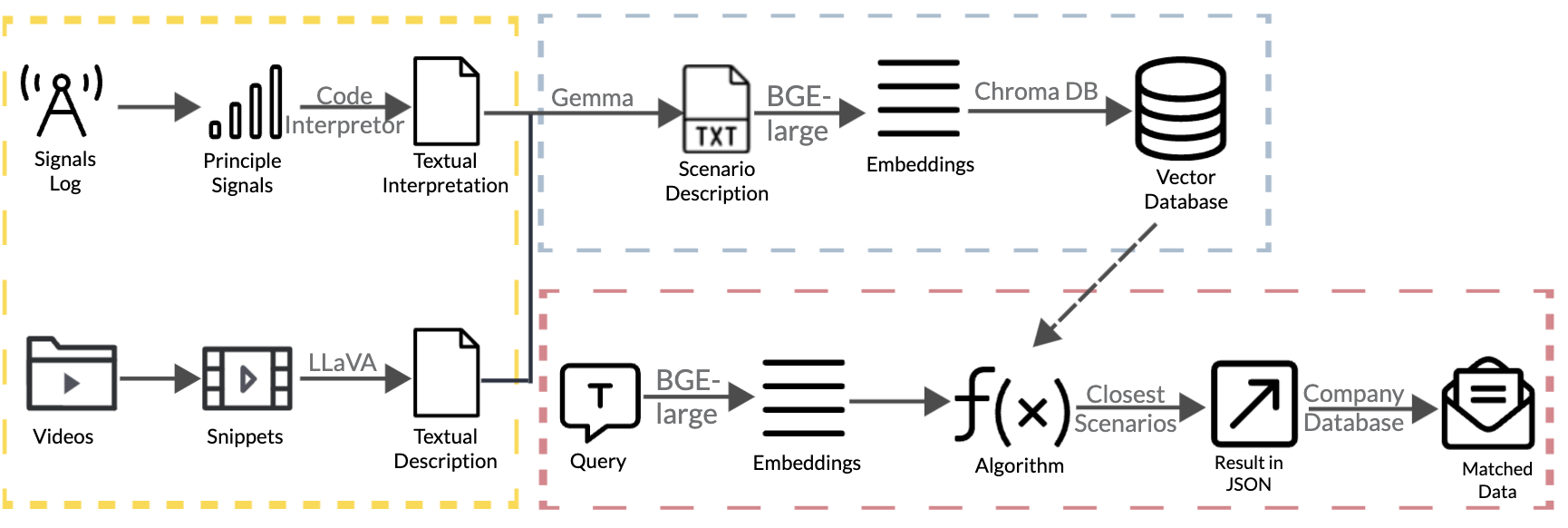}
    \caption{The data retrieval pipeline consists of three stages: converting signal and video logs into text (yellow), combining and embedding these descriptions and storing the embeddings (grey), and retrieving the most similar scenarios to a natural language query (pink)}
    \label{fig:model}
\end{figure}

The architecture we use is shown in Figure \ref{fig:model}, it is a data retrieval pipeline with the following stages:
\begin{itemize}
    \item Preprocessing (yellow box): converts signal logs and video logs into textual descriptions.
    \item Combining Descriptions and Embedding (grey box): combines the textual descriptions via text generation model(Gemma 7B), extracts embeddings(via BGE-large) and stores them in the form of vector through ChromaDB.
    \item Querying Specific Scenarios (pink box): queries specific scenarios using natural language. It compares the similarity between query embeddings and saved embeddings via ChromaDB, then displays the top N closest scenarios based on the settings. The results are returned in a JSON file containing metadata and a link to the Zenseact's internal database, which can be used to access the matched scenarios.
\end{itemize}

\subsection{Models}
Each of the three parts of the architecture use distinct models due to different modalities of the data in each stage.  

\textbf{LLaVA-1.5-7b:} 
The Large Language-and-Vision Assistant (LLaVA) is a multimodal language model that excels across various benchmarks by utilizing GPT-4 generated language-image instructions during training. We opted for the smallest 7-billion-parameter model due to its capacity to handle prompts up to 4,096 tokens and process images at a resolution of 336x336 pixels \cite{liu2024visual}. 

\textbf{Gemma-7b:}
Gemma, an open-source model family from Google \cite{team2024gemma}, offers two sizes: 2 billion and 7 billion parameters. After assessing both, Gemma 7b emerged as the preferred choice. While the 7b model utilizes 4-bit quantization for its weights, the 2b model employs 8-bit quantization. Despite the higher resolution of weights in the 2b model, the advantages of the larger model size outweighed this consideration. With a token limit of 8,192, ample space is available for few-shot prompting and generating lengthy descriptions; however, the quantization might potentially hinder the model's ability to comprehend extensive inputs and outputs.

\textbf{BGE-large:}
To facilitate semantic comparison between generated scenario descriptions and user queries, we employ a text embedding model. After evaluating various options using the Hugging Face Massive Text Embedding Benchmark (MTEB) Leaderboard \cite{muennighoff2022mteb}, we prioritize the "retrieval" performance, assessing how effectively a model can compare a query with a collection of documents and retrieve the most relevant ones. A standout performer in this metric is BGE-large, boasting high scores while maintaining a relatively small size of 335 million parameters. BGE is a part of C-pack, a comprehensive package comprising training data, benchmarks, and embedding models. Available in different sizes, we opt for the largest model due to its superior performance, as indicated in \emph{C-pack: Packaged resources to advance general chinese embedding.} \cite{xiao2023c}.

\textbf{Vector database (ChromaDB)}
After the scenarios have been processed and embedded, we store them in a vector database. For this, we have chosen to use ChromaDB since it is an open-source solution that is easy to implement. ChromaDB creates a local database, called a collection where the scenario descriptions can be stored with their respective embedding, and it also supports specifying metadata for each entry. We utilized metadata to store what vehicle, log file and the time of the scenario, this is useful when retrieving the scenarios since it makes it easier for users to find more data for the corresponding scenario. ChromaDB integrates with the local embedding model, when a scenario is added to the collection, an embedding is automatically created using the model and then stored in the collection. When you query for a scenario in the collection, ChromaDB also handles the vector similarity search, we chose to use squared Euclidean distance as the distance metric between embedded scenarios.

\section{Evaluation}

\subsection{Genius Interface}
\begin{figure}
    \centering
    \includegraphics[scale=0.4]{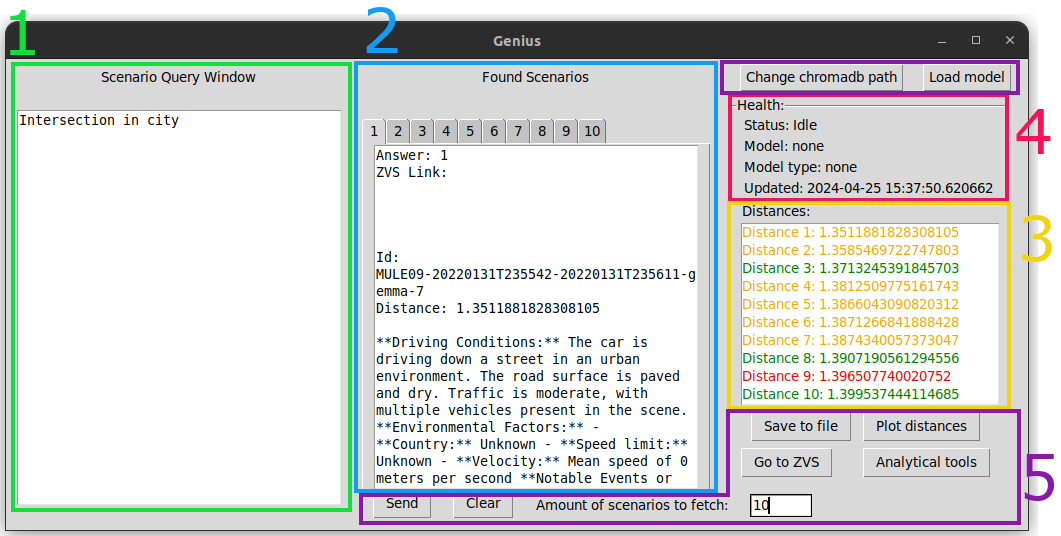}
    \caption{User interface for scenario querying.}
    \label{fig:genius_interface_artifact}
\end{figure}

To evaluating the generated descriptions and embeddings, we aim to provide an interface that enables users to query our dataset and visualize the results. The interface for 'Genius' streamlines the querying and debugging processes for users. Illustrated in Figure \ref{fig:genius_interface_artifact}, the interface is divided into distinct sections.

\textbf{Scenario Query Window:} This section enables users to input their query using natural language descriptions of scenarios.

\textbf{Found Scenarios Window:}
Here, extracted scenarios resulting from the query are displayed in ascending order in tabs. Each tab presents the scenario description alongside relevant metadata such as timestamp, ID, distance, and a link to a visualization tool provided by the Zenseact. 

\textbf{Distance List Window:}
This section lists the distances for each extracted scenario in ascending order. 

\textbf{Status Window:}
A status window provides users with real-time updates on the API's operational state and health.

\textbf{Miscellaneous Buttons:}
Utility buttons and an entry field enhance user experience. These include a 'Send' button for query submission (with 'Enter' key mapping for convenience), a 'Clear' button to reset the interface, and analytical tools for evaluating search results. 

\subsection{Model Selection}

To quantitatively evaluate the impact of the chosen text generation model on the final extraction, we designed a test process. In this process, we query Genius for a scenario known to exist within a predefined set of scenarios. We then group the distances of each response based on the originating scenario and plot these distances. This approach illustrates the variation within the same scenario and highlights the differences between correct and incorrect answers.

Our test set includes descriptions of each scenario, with 10 iterations totaling 80 embedded scenarios. Each iteration yields different results, enabling us to test the variability and stability of the process. This procedure is repeated for each model to ensure consistent evaluation.

\begin{table}[htbp]
    \caption{Table of comparison metrics for each model. The best-performing metric for each category is underlined.}
    \centering
    \resizebox{\textwidth}{!}{
        \begin{tabular}{lcccc}
            \toprule
            & Mistral 7b & Gemma 2b & Gemma 7b & Starling 7b \\
            \midrule
            Mean Distance to Correct Answers& 1.093&   1.061 & 1.081& 1.083 \\
            Mean Distance to Incorrect Answers& 1.433& 1.411& 1.419 & 1.417\\
            Mean Distance Difference & 0.340 & 0.350 &0.338 & 0.334\\
            Highest Distance to Correct Answer& 1.147 & 1.137 &1.133 & 1.139\\
            Lowest Distance to Incorrect Answer& 1.303 & 1.289 & 1.309& 1.289\\
            Smallest Distance Difference& 0.156& 0.152 & 0.176& 0.150\\
            Average Standard Deviation of Scenarios & 0.033 & 0.041 & \underline{0.028} &0.038\\
            \bottomrule
        \end{tabular}
    }
    \label{table:model-eval_table}
\end{table}

\begin{figure}
    \centering
    \includegraphics[scale=0.4]{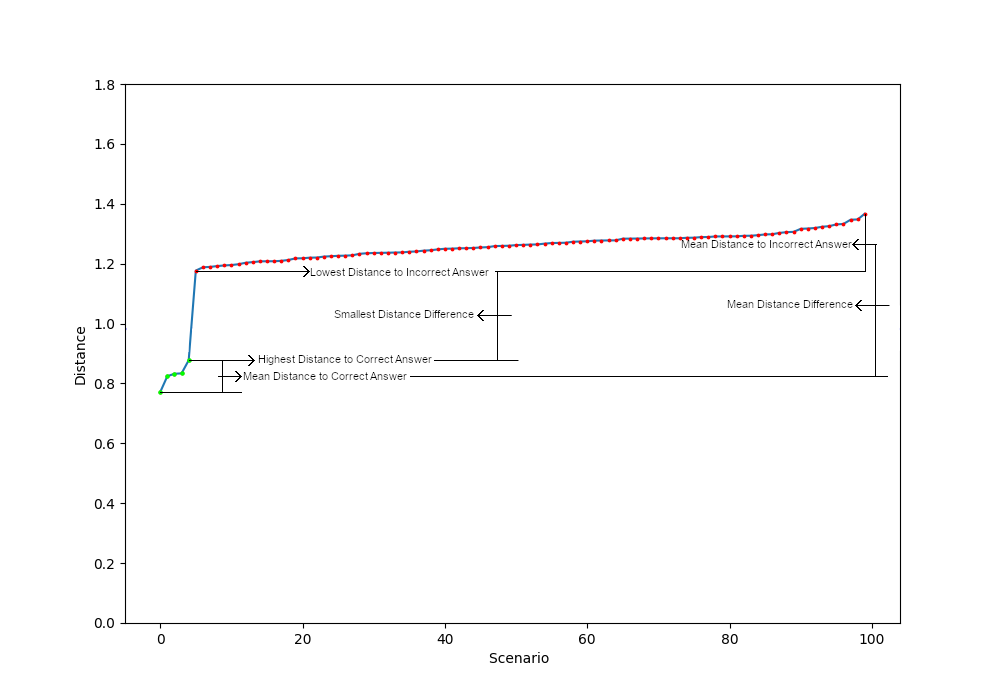}
    \caption{Graph exemplifying the different metrics used.}
    \label{fig:metrics}
\end{figure}

Table \ref{table:model-eval_table} and Fig \ref{fig:metrics} shows the comparison metrics used to evaluate the models. Emphasis should be placed upon the metrics \textit{Smallest Distance Difference} as well as \textit{Average Standard Deviation of Scenarios}, as they show the distinction between correct versus incorrect answers and the stability of the answers respectively. Metrics where both outliers are considered and excluded are shown.

Looking at the results Gemma 7b shows by far the best results, outperforming the others in most metrics, most importantly in both \textit{Smallest Distance Difference} and \textit{Average Standard Deviation of Scenarios}.

\subsection{Retrieval Evaluation}


To quantitatively evaluate the retrieved embedding for the 100 scenarios, a few varied queries are analyzed using the following metrics:
\begin{itemize}
    \item Largest Gap (LG): Measures the largest gap in distance between adjacent scenarios
    \item Min Distance (Min D): Measures the distance to the closest scenario
    \item Max Distance (Max D): Measures the distance to the furthest scenario
    \item Range: Measures the range of distances, e.g. Max Distance - Min Distance
    \item Standard Deviation (SD): Measures the standard deviation of the scenario distances
    \item Relative Largest Gap (Rel. LG): Measures the Largest Gap relative to the Range
\end{itemize}

\begin{table}[h]
    \caption{Measured Results for Different Scenarios(\romannumeral 1. Snowy highway in Sweden, \romannumeral 2. Driving under bridge, \romannumeral 3. Car Accident, \romannumeral 4. Car Changed Lane) and Average Across Scenarios}
    \centering
    \begin{tabular}{ccccc|ccc}
    \hline
    \multicolumn{1}{l}{} & \multicolumn{4}{c|}{Scenarios}                          & \multicolumn{3}{c}{Average Across Scenarios}                      \\ \hline
    & \romannumeral 1     & \romannumeral 2 & \romannumeral 3 & \romannumeral 4        &   Correct Set    &   Incorrect  Set  & Test Set\\ \hline
    LP                    & 0.09962       & 0.2996       & 0.0201   & 0.009947    & 0.09903      & 0.0246         & 0.046            \\
    Min D                 & 1.226         & 0.7712       & 1.155    & 1.355       & 1.113        & 1.248          & 0.8685           \\
    Max D                 & 1.512         & 1.367        & 1.343    & 1.519       & 1.476        & 1.470          & 1.185            \\
    Range                 & 0.2858        & 0.5956       & 0.1879   & 0.1640      & 0.3421       & 0.2226         & 0.3168           \\
    SD                    & 0.03688       & 0.1031       & 0.03477  & 0.03494     & 0.05973      & 0.04446        & 0.06401          \\
    Rel. LG               & 34.9\%        & 50.3\%       & 10.7\%   & 6.1\%       & 26.6\%       & 11.1\%         & 14.5\%           \\ \hline
    \end{tabular}
    \label{table:measured_results}
\end{table}

In order to gain an metric to evaluate correct from incorrect answers an Average Relative Largest Gap (ARLG) metric is calculated. By querying a set of queries with known correct answers and taking the ARLG, a standard ARLG for correct answers can be calculated.
By comparing the ARLG measured for the test set, 14.5\% against the baseline ARLG $(26.6\% + 11.1\%)/2 = 18.8\%$, it is observable to see that the average characteristics of the test set answers are closer to the characteristics of a query without answers rather than one with correct answers. As the test queries should all contain answers, this is not a desirable outcome. However, this is not due to a poor search functionality of the solution as by manually checking we can see that the closest answers given are indeed either the scenario previously analyzed or a similar scenario that fits the query. However, the issue lies in when queries are too broad, when a query fits for a large share of the scenarios in the dataset, the ARLG will be low and the curve looks similar for a query with no correct answers even though a large number of answers are in fact correct.

If attention is instead diverted to the Z-score validation, this metric reports that 8 of the 10 test queries have correct answers. While this is a much better result and it accurately confirms large majority of the set, it falls prey to the same issue shown by the ARLG, which is where the query is too general and contains too many answers which offsets the whole curve and decreases the chances for outliers.

\begin{figure}[h!]
    \begin{minipage}{0.5\textwidth}
        \centering
        \includegraphics[width=\linewidth]{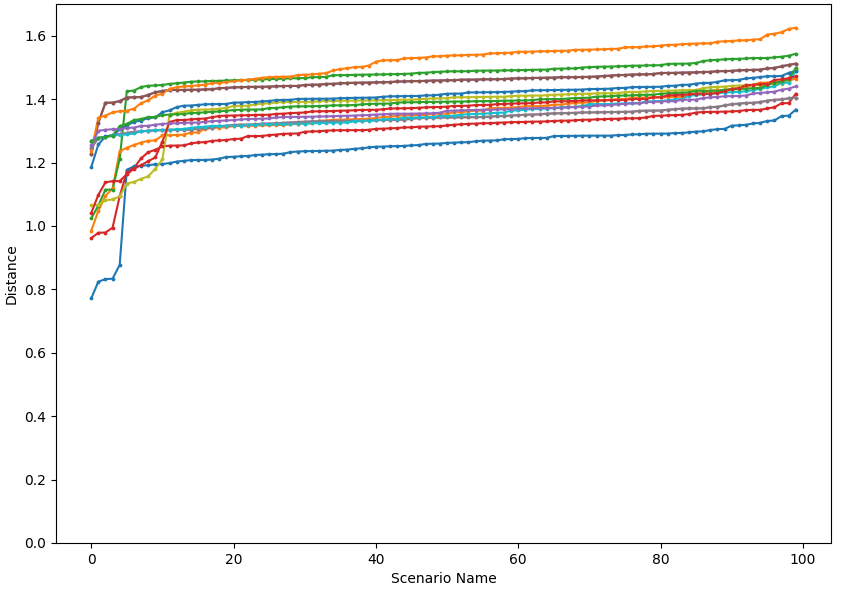}
        \captionsetup{justification=centering} 
        \caption{All scenario distances for the set of correct queries}
        \label{fig:distance_ave}
    \end{minipage}%
    \begin{minipage}{0.5\textwidth}
        \centering
        \includegraphics[width=\linewidth]{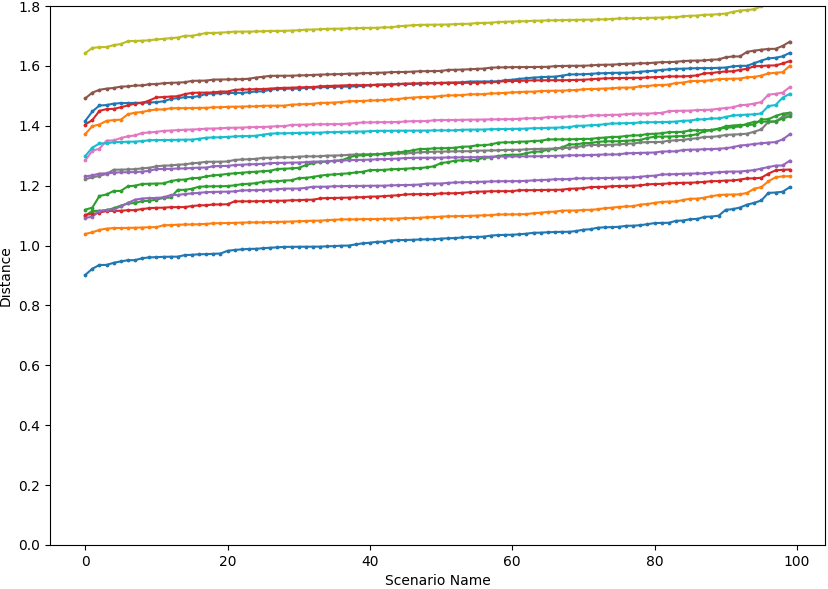}
        \captionsetup{justification=centering} 
        \caption{All scenario distances for the set of incorrect queries.}
        \label{fig:distance_all_incorrect}
    \end{minipage}%
\end{figure}

We plot the distances between the query scenario and the rest of the scenarios, with each line representing a different query. The correct and incorrect sets are plotted separately. Figure \ref{fig:distance_ave} shows that the correct answers exhibit distinctly lower distances, while the incorrect answers form a slowly ascending line with relatively uniform distances. There is a clear distinction between correct and incorrect answers, as evidenced by the plot and the Largest Gap metric. Conversely, for the incorrect set in Figure \ref{fig:distance_all_incorrect}, there is not a clear observable distinction; the curve of the plot demonstrates a slowly ascending line, indicating no clear separation between any of the answers, which is also reflected in the Largest Gap metric.

\subsection{Survey}

To evaluate the quality of the scenario descriptions generated, a form was sent out to engineers working with the log data at the Zenseact. The 8 scenarios were split into 4 forms with 2 scenarios in each. The forms included links to an internal visualization tool for the specific scenarios along with the corresponding descriptions generated by Genius. The developers were asked to grade the description and point out any areas of improvement that could be worked on along with any factual errors in the description.

The 16 responses resulted in a mean value of 3.3125, which according to our rank of satisfaction is "Acceptable description of the scenario, capturing some of the important aspects.". Out of the eight scenarios there was only one scenario that had a score under 3, the same scenario was rated twice with the score 1.

\subsection{Evaluation of RQs}
\noindent

\textbf{RQ1:} 
Current LLM models can not fully utilize the vast tabular data that was used in this paper, at the very least not with the smaller size of models that we deployed, instead, some sacrifice will have to be made in what signals the model should utilize. Based on our findings from evaluating the descriptions generated when using a subset of signals, the model can generate descriptions that are acceptable and capture some of the most important aspects of a vehicle scenario.

\textbf{RQ2:}
During embedding and model evaluations, we found that using both signals and video information as input resource does indeed enhance the descriptions of the scenarios. The model effectively captures key aspects such as weather, road, and traffic conditions, and type of environment. Combining these visual data with signal data, which capture more complex car experiences, results in a more detailed and complete scenario description. However, it sometimes misses details like other cars and confuses urban with rural settings, likely due to heavy quantization or the model's 336x336 pixel resolution. Additionally, the multimodal LLM processes images one at a time; while it is feasible to input images sequentially to denote a video, we opted for a single frame per scenario to expedite processing, albeit at the expense of potentially missing subsequent scenario details.

\textbf{RQ3:}
Our finding shows that using an embedding model to embed the semantic meaning of a vehicle scenario works well to group similar scenarios close together.
Some limitations exist in our solution, the text embeddings model good semantic representation of numbers makes querying for number specific scenarios, such as specific speeds very difficult. Another limitation is that similar wording that mean different things will embed close to each other in vector space, making it hard to find scenarios that are not very specific, for example "change lane" will be close to "driving in the left \textbf{lane} with an upcoming \textbf{change} in speed limit".
The solution shows the best result when the query is specific enough that only a few of the scenarios in the data set can fulfill the query.

\section{Conclusion and Future Research}

The current implementation of the artifact may not replace SQL-based scenario searches but could serve as a complementary tool for finding specific scenarios challenging to detect through signal data alone, such as "snowy scenario in a rural area" or "intersection in an urban area with traffic lights," accessible only from video data.

The artifact's search functionality has proven stable, yielding satisfactory results. When querying for existing signals and scenarios, similar scenarios are typically returned first, albeit not always in the most fitting order. Searching for scenarios functions well at a larger scale, although finding homogeneous scenarios like "lane change" remains challenging due to its frequent occurrence across various scenarios.

While determining answer correctness without manual verification was challenging, the keyword search method effectively flagged incorrect answers by identifying semantic similarities. However, it struggled with distinguishing between correct and incorrect answers when common keywords were present in all scenarios.

The biggest limitation with the current approach is that it can not process all the signals in the data, this is due to the fact that there are several thousand signals and LLMs have a limited context size. Due to hardware limitations we also choose to process just one frame of the camera feed for each scenario, this results in limited knowledge of information that is only present in the camera feed.

The set of scenarios that were generated for the evaluation survey were specifically chosen to contain events that are captured in these signals, therefore the results might not have been as clear with a different selection of scenario categories. However, information from video footage of the scenario was also utilized, therefore information that was present in a signal, for example if the car was in a tunnel or not, could have been captured anyway even that signal was missing. This fact might have impacted the results of the evaluation form that we sent out to developers at Zenseact in our favor, since we are using scenarios where the utilized signals are relevant.

To improve results, future work should focus on three main areas: first, filter and preprocess signal data by incorporating more relevant signals and excluding those that remain unchanged during scenarios; second, increase the number of frames used from video footage to generate more detailed descriptions, or shorten scenario lengths to include more frames; and third, train a customized LLM, either by developing separate models to process signal data and generate descriptions or by creating a multi-modal model capable of handling both inputs.

\begin{credits}
\subsubsection{\ackname} We would like to thank Zenseact for giving us the opportunity to conduct our project at their office and providing us with all the required resources.

This work has also been partially funded by Software Center, a collaboration between University of Gothenburg, Chalmers and 18 universities and companies -- \url{www.software-center.se}.
\end{credits}

%
%
\bibliographystyle{splncs04}
\bibliography{reference}

\end{document}